\documentclass{cargese}\usepackage{graphicx}
\usepackage{amsmath,amssymb}

\newif\ifpreprint
\preprinttrue

\let\footnote\savefootnote
\let\footnotetext\savefootnotetext 
\let\footnoterule\savefootnoterule 
\normallatexbib

\newcommand\adst{{$AdS_2$/CFT$_1$\index{$AdS$/CFT!$AdS_2$/CFT$_1$}}}

\newcommand\skipthis[1]{{}}

\begin{document}


\articletitle[Anti-de Sitter Fragmentation]{Anti-de Sitter Fragmentation%
\ifpreprint\footnote{Gong Show talk given at the Cargese '99 ASI ``Progress
in String Theory and M-Theory''}\fi}


\author{Jeremy Michelson}


\affil{New High Energy Theory Center \\ Rutgers University \\ 
       126 Frelinghuysen Road \\ Piscataway, NJ~~08854 \\ USA }
%
\email{jeremy@physics.rutgers.edu }
%

\ifpreprint 
\begin{abstract}
Some general aspects of the \adst\ correspondence, previously discussed in
hep-th/9812073, are summarized.  The
majority of this summary is devoted to the question of where the CFT$_1$
should live.  Almost as a byproduct, the decay---or
fragmentation---of $AdS_2$ is also discussed.
\end{abstract}
\fi


\ifpreprint \else \section*{} \fi

As is hopefully symptomatic of the number of contributions in 
\ifpreprint the Cargese '99 \else these \fi
proceedings devoted to this conjecture, the $AdS$/CFT
correspondence\index{$AdS$/CFT}
has been very fruitful in illuminating aspects of both large $N$ gauge
theory, and supergravity on $AdS$ backgrounds.  The $AdS_5$/CFT$_4$
conjecture is both very interesting and widely studied, because it sheds
light on four-dimensional gauge theories.\index{$AdS$/CFT!$AdS_5$/CFT$_4$}%
\footnote{The reader may have already noticed a dearth of references in
this summary.  That is because we are making every effort to acknowledge
only our own research; all other citations are completely inadvertent.}
The $AdS_3$/CFT$_2$\index{$AdS$/CFT!$AdS_3$/CFT$_2$} conjecture has
also been widely studied.   Because of the simplifications that result in
lower dimension---and because of the amount of knowledge that has
accumulated on two-dimensional CFTs---the
$AdS_3$/CFT$_2$\index{$AdS$/CFT!$AdS_3$/CFT$_2$}
correspondence has been very useful at elucidating information on $AdS_3$
supergravity.  For example, a detailed microscopic description of the BTZ
black hole\index{BTZ} has been obtained in this way\skipthis{~\cite{as}}.

Continuing down in dimension, the 
\adst\ conjecture is important for the
understanding of four and five-dimensional black holes\index{black
hole}---the near horizon
geometry of such black holes is respectively $AdS_2\times S^2$ and
$AdS_2\times S^3$.\index{black hole}
One might think that this
lowest-dimensional correspondence should be the
simplest of all, and the easiest to check.  In fact, very little is known
about either side of the \adst\ correspondence.  In this summary, we
will discuss some aspects of the $AdS_2$\index{$AdS$!$AdS_2$} side of the
correspondence.  A
more complete discussion, and references, can be found in~\cite{us}.
For later work on this subject, we refer the reader in part
to~\cite{msas,jmms,corley,ll,jmas1,jmas2} and references therein.

\begin{figure}[t]
\begin{center}
\includegraphics[width=3in]{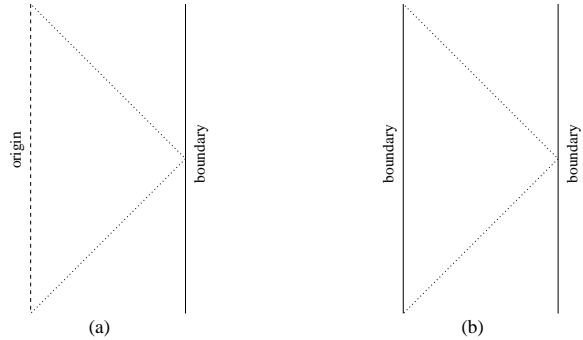}
\end{center}
\caption{The Penrose diagram\index{$AdS$!Penrose diagram} for (a)~%
$AdS_{D\geq3}$ (each point is an $S^{D-2}$) and (b)~$AdS_2$.  The solid
line is the boundary, the dotted
lines delimit the Poincar\'{e} patches, and the dashed line indicates the
origin ($r=0$) of the global coordinates~\eqref{adsds}.}\label{fig:bads}
\end{figure}

\begin{figure}[b]
\begin{center}
\includegraphics[width=3.5in]{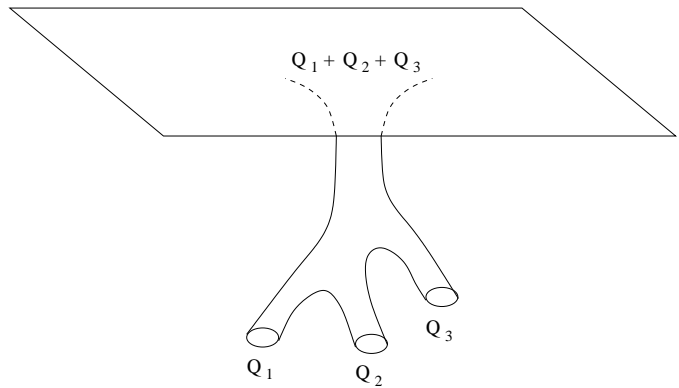}
\end{center}
\caption{Approaching the near-horizon limit of the multi-black hole 
 configuration
considered.  From afar, the geometry looks like that of a single black hole
but as the throat is approached it splits up into several (three depicted) black
holes.  In the near-horizon limit, only the (four) throats
remain.}\label{fig:nhmbh}
\end{figure}

One of the main puzzles in understanding the \adst\ correspondence, is the
question of where the CFT should ``live''.  In higher dimensions,
the CFT is usually taken to live on the boundary of the $AdS$\index{$AdS$} space;
however, $AdS_2$\index{$AdS$!$AdS_2$} has {\em two} boundaries, as shown in
figure~\ref{fig:bads}.
To be precise,
in global coordinates\index{$AdS$!global coordinates},%
\footnote{In Poincar\'{e} coordinates\index{$AdS$!Poincar\'{e} coordinates}
($0<z<\infty$, $x^\mu \in \mathbb{R}^{D-2,1}$)---%
which cover only a patch of $AdS_D$---the metric is
$ds^2 = ({dz^2 + \eta^{(D-1)}_{\mu \nu} dx^\mu dx^\nu})/{z^2}$.}
the $AdS_{D\geq3}$ metric takes the form
\begin{equation} \label{adsds}
ds^2 = \cosh^2 r\, dt^2 + dr^2 + \sinh^2 r \, d\Omega_{D-2}^2,
\end{equation}
where $0<r<\infty$ and the boundary is the $S^{D-2}\times\mathbb{R}$ at
$r=\infty$.  If $D=2$, then the last term of equation~\eqref{adsds} is
absent, $-\infty<r<\infty$, and the boundaries are at $r=\pm \infty$.
So, are there {\em two} CFTs and if not, then on which boundary does the
CFT live?

A strong hint that the CFT lives on only one boundary is given by the
following consideration.  We can obtain $AdS_2\times S^2$ from the
near-horizon geometry of a four-dimensional extremal Reissner-Nordstr\"{o}m
black hole\index{black hole!Reissner-Nordstr\"{o}m}.  In fact, we can
obtain a much richer
system by considering the geometry obtained by taking many extremal black
holes and allowing them to approach each other as the near-horizon limit is
taken---see figure~\ref{fig:nhmbh}.

\begin{figure}[t]
\begin{center}
\includegraphics[width=.6in]{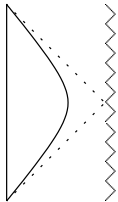}
\end{center}
\caption{The Penrose diagram for a small black hole spherically averaged
about a large black hole.  The curve is the trajectory of the small black
hole.} \label{fig:sing}
\end{figure}

In particular, we examine the system with
two black holes with charges $Q_{1,2}$ and consider the case when $Q_2\ll
Q_1$.  Then, we can recover an $AdS_2 \times S^2$ geometry by spherically
averaging the small, ``test'' black hole about the large black hole.
More precisely, we obtain an asymptotically $AdS_2\times S^2$ geometry in
this way.  In the two-dimensional theory, the second black hole appears as
a test particle, and, as depicted in figure~\ref{fig:sing}, one of the
boundaries---namely the one just inside the horizon of the large black
hole---of the spacetime has become singular.%
\footnote{A derivation can be found in ~\cite{us}.}
Thus, we conclude that the CFT lives only on one boundary.

Finally---and to justify the title of this summary---we note that the analytic
continuation of this story to Euclidean signature, is an instanton
description of $AdS_2$\index{$AdS$!$AdS_2$} fragmentation or
splitting\index{$AdS$!fragmentation}.  Roughly,
the throat
at the top of figure~\ref{fig:nhmbh} is the ``initial'' state---a large
$AdS_2 \times S^2$---which then splits into many smaller throats.  

One
might ask whether higher-dimensional $AdS$\index{$AdS$} spaces can also split.  The
answer turns out to be negative: the action for the
$AdS_2$\index{$AdS$!$AdS_2$} instanton is
precisely one-half the change in the entropy (so that the rate of
black hole splitting is suppressed by the entropy) but for
higher-dimensional
splitting, there is an infinite factor, related to the fact that the
geometry is the near-horizon limit of a noncompact $(D-2)$-brane.
Thus, only $AdS_2$\index{$AdS$!$AdS_2$}
fragments.%
\footnote{A detailed alternative explanation can be found in~\cite{sw}.}

One might also ask how $AdS_2$ fragmentation\index{$AdS$!fragmentation}
manifests itself in the CFT on
the boundary.  This is still an open question.


\begin{acknowledgments}
I thank J.\ Maldacena and A.\ Strominger for very fruitful collaboration on
this topic, and the organizers of this excellent school for the opportunity
to attend.  This work was supported by an NSF Graduate
Fellowship, an NSERC PGS B Scholarship
and DOE grant DE-FGO2-91ER40654.
\end{acknowledgments}



%

\end{document}